\documentclass[prl,twocolumn]{revtex4}
\usepackage{amsfonts}
\usepackage{amsmath}

\begin{document}

\title{Quantum Statistics with Classical Particles}
\author{Daniel Gottesman}
\email[E-mail: ]{dgottesman@perimeterinstitute.ca}
\affiliation{Perimeter Institute for Theoretical Physics,
Waterloo, ON N2L 2Y5, Canada}

\begin{abstract}
Indistinguishability of particles is normally considered to be an
inherently quantum property which cannot be possessed by a
classical theory.  However, Saunders~\cite{saunders} has argued
that this is incorrect, and that classically indistinguishable
particles are possible.  I make this suggestion concrete by
describing a class of microscopic classical theories involving
indistinguishable particles hopping stochastically on a graph, and
show that it should be possible to experimentally create a
physical system realizing a simple model by continuously observing
atoms trapped in an optical lattice.  The indistinguishable
classical particles obey Bose-Einstein statistics, display the
associated clustering phenomena, and in appropriate models, can
even undergo Bose-Einstein condensation.
\end{abstract}

\maketitle

Since the discovery of quantum mechanics, much has been made of
phenomena such as lasers, Bose-Einstein condensation,
superfluidity, and superconductivity. Theoretical and experimental
studies of these phenomena have revealed a variety of fascinating
and counter-intuitive behaviors, consequences of the deviations
from classical Maxwell-Boltzman statistics displayed by quantum
particles.  Indeed, they are associated so strongly with quantum
mechanics that they are usually described as ``macroscopic quantum
phenomena.'' Indistinguishability, we are told, is a property that
can only be possessed by quantum particles and not by classical
ones.

Given the importance of these macroscopic quantum behaviors, it is
worthwhile examining this belief carefully.  While all our known
fundamental quantum particles are either bosons or fermions,
certainly it is possible in principle to have quantum particles
which are distinguishable, perhaps because of some internal
quantum number which can take a large variety of different values.
But is it possible to have classical bosons or fermions?

The standard folklore of physics says ``no.''
Bach~\cite{trajectories}, for instance, has argued that a theory
with indistinguishable classical particles is impossible because
the trajectories would distinguish those particles. However,
Saunders~\cite{saunders} has recently argued that one {\em can}
imagine classical indistinguishable particles. Rather than having
a separate configuration space for each particle, the space of
configurations would be a collective one, indicating only the
number of particles present at each location. In Saunders' models,
these indistinguishable classical particles still effectively
obeyed Maxwell-Boltzman statistics. The difference with quantum
particles, he says, is that quantum particles also experience a
concentration of measure to a lattice of points (in a system of
finite size), whereas classical particles can be anywhere.

However, classical particles which are hopping from point to point
on a discrete lattice do have their probability measures
concentrated on points.  In this paper, I give explicit
microscopic stochastic models with indistinguishable classical
particles hopping on a discrete lattice.  The particles are
classical in the sense that at all times, they have definite
positions.  There are no superpositions and interference cannot
occur. These models display Bose-Einstein statistics, unlike
Saunders' continuous models, and can even display Bose-Einstein
condensation, normally considered a purely quantum phenomenon.
% Presumably a variety of other
% macroscopic quantum phenomena are possible too.
Saunders also did not specify the microscopic dynamics of his
particles, leaving open the possibility that a contradiction
arises somewhere, whereas my models have a complete microscopic
description.

The difference between classical indistinguishable particles in
discrete space and continuous space can be easily understood ---
indistinguishability only has an effect on the equilibrium state
of a system when particles have a good chance to be in the same
state.  For classical particles in a continuous space, the
probability is $0$ of two particles having the same position and
momentum, whereas quantum particles have a finite probability of
overlapping due to the uncertainty in position and momentum. On a
discrete configuration space, however, classical particles can
arrive at the same state with finite probability, and thus can
display the distinctive behaviors normally associated with quantum
indistinguishability.

Furthermore, the models that I present of classical
indistinguishable particles are not just theoretical fantasies; it
should be possible to realize some simple instances
experimentally.  In particular, spin-$0$ atoms confined to an
optical lattice satisfy the Bose-Hubbard model to a good
approximation~\cite{jaksch}. This produces a system of bosons
hopping in a discrete configuration space.  These particles are
quantum, of course, not classical, but suppose we observe the
system at intervals of length $\Delta t$ to see how many atoms are
present at each lattice site.  At each observation, the system
collapses to some definite classical configuration.  The required
measurement non-destructively localizes all particles to
individual wells in the optical lattice.  This, unfortunately, is
beyond current experimental techniques, but may not be for long.
Between observations, some particles may hop from one site to
another, but if we observe frequently enough, it is extremely
unlikely that more than one particle will hop in time $\Delta t$.
Interference between different paths is then impossible, and the
system becomes, to an excellent approximation, a system of {\em
classical} bosons.

After I completed the research reported here, I discovered similar
work by Kaniadakis and Quarati~\cite{cb,cb2}.  Our models are
related but distinct.  Their models are primarily built in
continuous systems with an explicit smearing function in space and
velocity to allow the classical particles to overlap (although
they do also give one example on a lattice).  Another similar
approach is the quantum Boltzmann master equation by Jaksch et
al.~\cite{QBME}, which was used to numerically simulate
Bose-Einstein condensation. I work exclusively in discrete
systems, allowing my models to be simpler than either of these and
to more clearly separate quantum mechanical behavior from
indistinguishability; indeed, Kaniadakis and Quarati describe
indistinguishability in their models as a residual quantum effect
rather than a separate phenomenon, and Jaksch et al.\ make no
comment at all about classicality.

Let us be concrete.  Suppose we have some graph $G$.  We can
associate to it a simple model of classical bosons by defining the
state of the system to be a function $n: G \rightarrow \mathbb{N}$
which gives the number of particles $n_i$ at each site $i$ in the
graph.  The total number of particles is $N = \sum_i n_i$.  These
particles can hop from site to site, conserving the total number.
For simplicity, we impose artificially the constraint that only
one particle in the whole system can hop in a single time step.
This means the particles will, strictly speaking, be interacting,
but if we work in the limit where hopping is a rare event, then
the interaction is negligible.

Suppose $i$ and $j$ are two adjacent sites.  Then let us assume
there is a probability $p$ to go in one time step from a
configuration $(1_i, 0_j)$ (with $1$ particle at site $i$ and $0$
particles at site $j$) to $(0_i, 1_j)$ (with $0$ particles at $i$
and $1$ particle at $j$).  This probability is independent of $i$
and $j$ (provided they are adjacent) and is the same regardless of
the number of particles at all other locations.

However, it is not immediately clear from this what the
probability should be of hopping from the configuration $(c_i,
d_j)$ to $((c-1)_i, (d+1)_j)$.  One constraint that seems sensible
is to require that the probability of hopping from configuration
$A$ to configuration $B$ is the same as the probability of hopping
from configuration $B$ to configuration $A$.  The analogous system
of distinguishable classical particles hopping on the graph $G$
has this property, since the probability of a {\em particular}
particle hopping forward is the same as its probability of hopping
backwards. This property also corresponds more or less to having a
Hermitian Hamiltonian in the analogous quantum system.

If we add the condition that the probability of hopping out of
site $i$ is proportional to the number of particles at $i$, we get
the probability rule
\begin{equation}
P\left[(c_i, d_j) \mapsto ((c-1)_i, (d+1)_j)\right] = c(d+1)p,
\label{eq:rule}
\end{equation}
again assuming the lattice sites $i$ and $j$ are adjacent.  We
should add the additional constraint that $p$ is small, so that
the total probability of hopping is always less than one. Then for
any configuration $A$, we can calculate the probability of staying
at the same configuration as $1-\sum_B P[A \mapsto B]$, where the
sum is taken over all configurations $B$ which are one hop away
from $A$.  With $N$ particles, it is sufficient to take $[N^2/4 +
(g + 1/2)N] p < 1$, where $g$ is the maximum degree of any node in
$G$.  Systems with distinguishable particles can also obey this
transition rule, leading to similar effects (e.g.,~\cite{BB}).

This prescription has the virtue that it agrees with the system
produced when we frequently observe particles obeying a
Bose-Hubbard model, as with spin-$0$ atoms in an optical lattice.
The standard optical lattice is a cubic lattice, but it is also
possible to make other graphs by blocking some sites using a
Fermi-Bose mixture~\cite{FBmix}. The Bose-Hubbard model~\cite{BH}
has the Hamiltonian
\begin{equation}
H = -J \sum_{\langle i,j \rangle} b_i^\dagger b_j + \sum_i
\epsilon_i \hat{n}_i + (U/2) \sum_{i} \hat{n}_i (\hat{n}_i - 1),
\end{equation}
where $b_i$ is the annihilation operator at site $i$ and
$\hat{n}_i = b_i^\dagger b_i$ is the number operator at site $i$.
The basis states of this quantum system are the possible
configurations of the above model of classical bosons with the
same graph of possible sites.  If we observe the system at two
times separated by an interval of $\Delta t$, the probability of
hopping from a basis configuration $A$ to $B$ is
\begin{equation}
P(A \mapsto B) = \left|\langle A | e^{iH\, \Delta t/\hbar} | B
\rangle \right|^2.
\end{equation}
When $\Delta t$ is small, we can approximate
\begin{equation}
e^{iH\, \Delta t/\hbar} = 1 + iH\, \Delta t/\hbar + O((\Delta
t)^2).
\end{equation}
The $\hat{n}_i$ terms do not cause transitions, so for $A \neq B$,
we only need to consider the term $-J \sum_{\langle i,j \rangle}
b_i^\dagger b_j$ in $H$.  If $A$ and $B$ differ on the adjacent
sites $i$ and $j$, with $A$ including $(c_i, d_j)$ and $B$
including $((c-1)_i, (d+1)_j)$, then
\begin{eqnarray}
P(A \mapsto B) & = & \left(\frac{J\, \Delta t}{\hbar} \right)^2
\left|\langle c_i, d_j | b_i^\dagger b_j | (c-1)_i, (d+1)_j
\rangle \right|^2 \nonumber
\\
& & + O((\Delta t)^3).
\end{eqnarray}
Recalling that $b_i |c_i \rangle = \sqrt{c} |(c-1)_i\rangle$, we
find
\begin{equation}
P(A \mapsto B) = c(d+1) (J\, \Delta t/\hbar)^2 + O((\Delta t)^3),
\end{equation}
which agrees with eq.~(\ref{eq:rule}) using $p = (J\, \Delta
t/\hbar)^2$.  This approximation is valid when $gNJ\, \Delta
t/\hbar$, $(N\, \Delta t/\hbar)\max (\epsilon_i)$, and $N^2 U\,
\Delta t/\hbar$ are all much less than $1$.

The classical boson model is a Markov processes, and can thus be
described via a transition matrix $M$ giving the probability to
hop between configurations. Because the probability of hopping
forward is the same as the probability of hopping backwards, the
transition matrix is symmetric, and is thus doubly stochastic
(each row and each column of $M$ sums to $1$). For an arbitrary
initial state, a Markov process asymptotes to an equilibrium
state, an eigenvector of $M$ with eigenvalue $+1$. For a doubly
stochastic matrix, the uniform mixture over all configurations is
a possible equilibrium state. If the graph $G$ is connected, the
Markov chain is irreducible, and if $p$ is small, there is a
non-zero probability of staying in the same configuration, so the
Markov chain is aperiodic.  When both of these are true, the
uniform mixture will be the {\em only} equilibrium
state~\cite{Markov}.

The equilibrium state of the Markov process therefore corresponds
to thermodynamic equilibrium. For instance, suppose we consider an
extremely simple case with two lattice sites and two particles.
For classical distinguishable particles, there are four
configurations, two with one particle on each site, and two with
both particles on the same site.  The equilibrium state thus has
probability $1/2$ of having both particles in the same location.
For the classical boson model, there are only three
configurations, the same two with both particles on the same site,
but only one with one particle on each site.  The equilibrium
state thus has probability $2/3$ of having both particles in the
same location, displaying the usual clustering effect associated
with Bose-Einstein statistics. Naturally, this clustering becomes
more prevalent in larger systems with a high density of particles.

% In
% general, if we have a lattice with $V$ sites and $N$ particles, we
% can calculate $\sigma_i^2 = \langle n_i^2 \rangle - \langle n_i
% \rangle ^2$, the variance in the number of particles at a given
% site $i$. For distinguishable particles and large $N$ and $V$, we
% find $\sigma_i^2 = N/V$, whereas for indistinguishable particles,
% we have $\sigma_i^2 = N^2/V^2 + N/V$.  When $V$ is much larger
% than $N$, the particle statistics make little difference, because
% two particles are unlikely to be in the same location; when $V$ is
% significantly smaller than $N$, the bosons tend to cluster
% relative to the distinguishable particles, resulting in greater
% variance in the number of particles at an individual site.

This simple model of classical bosons illustrates the basic point
--- that classical particles can be indistinguishable --- but
cannot display the more exciting properties associated with
indistinguishable quantum particles.  If we want to make classical
analogues of macroscopic quantum phenomena, we will need models
where the particles have additional properties.  As a first step,
let us introduce energy into the model.  Since we wish energy to
be conserved at each transition, the new model will have two types
of particles, which I will call ``atoms'' and ``photons.''  This
enables us to have both interesting dynamics and conservation of
energy by shifting energy back and forth between the atoms and the
photons.

Atoms will hop on a graph, just as in the simple model.  However,
now each site $i$ in the graph will have an integral value of
energy $E_i$ associated with it.  We assume that adjacent sites of
the graph have energy values that are within $\pm 1$ of each
other.  (The energies of adjacent sites may be the same.)  The
number of atoms is conserved in each transition.  We will assume
the atoms are classical bosons, so the possible atom
configurations are described by the number of atoms at each site.
% However, we could make a similar model with distinguishable atoms.

Photons have a graph of their own, and may hop on it also as in
the previous model.  The photons will be indistinguishable
classical particles too. When photons hop between different sites
on the photon graph, the number of photons is conserved, but
photons can be created or destroyed when an atom makes a
transition.  In particular, we associate a photon site $k$ with
each edge $\langle i,j \rangle$ of the atom graph for which $|E_i
- E_j| = 1$.  When an atom makes a transition along the edge
$\langle i,j \rangle$, a photon must be created or destroyed at
the site $k$ in order to conserve energy.  Each photon is
considered to have unit energy.

We thus have a variety of possible types of transitions in this
model, with transition probabilities given in
table~\ref{table:energymodel}.
\begin{table}
\centering
\begin{tabular}{lr}
Transition type & Probability \\ \hline %
$i \rightarrow j$, $E_i = E_j$ & $n_i(n_j+1)p$ \\
$i \rightarrow j$, $E_i = E_j + 1$, create $\gamma$ at $k$ &
$n_i(n_j+1)(m_k+1)p'$ \\
$i \rightarrow j$, $E_i = E_j - 1$, destroy $\gamma$ at $k$ &
$n_i(n_j+1)m_kp'$ \\
$i \rightarrow j$, $|E_i - E_j| > 1$ & $0$ \\
$k \rightarrow l$ & $m_k (m_l + 1) q$
\end{tabular}
\caption{Transition probabilities in the classical boson energy
model. Sites $i$ and $j$ are adjacent atom sites with energies
$E_i$ and $E_j$. Sites $k$ and $l$ are adjacent photon ($\gamma$)
sites, and if appropriate, photon site $k$ is associated to the
edge $\langle i,j \rangle$. There are $n_i$ and $n_j$ atoms at
sites $i$ and $j$ respectively, and $m_k$ and $m_l$ photons at
sites $k$ and $l$.} \label{table:energymodel}
\end{table}
As with the previous model, we only allow a single particle to
make a hopping transition in each time step, plus the possibility
that a single photon is also created or destroyed.  To determine
the transition probabilities, it seems reasonable to assume the
probability of absorbing a photon increases proportionally to the
number of photons available to be absorbed.  From this, it follows
that the requirement that the Markov chain be doubly stochastic
implies that the probability to emit a photon into a site with
$m_k$ photons must be proportional to $m_k + 1$.  We allow
different probabilities $p$ and $p'$ for an atom hopping to a site
with the same or different energy, and a separate probability $q$
for photons hopping.

Assuming the atom and photon graphs are connected, and that $p$,
$p'$, and $q$ are sufficiently small, there is again a unique
equilibrium state, which is a uniform mixture over all
configurations.  There are two conserved quantities, the total
energy $E$ and the total number of atoms $N_A$.  If there are
$n_i$ atoms at site $i$, then the total number of photons must be
$n_P = E - \sum_i n_i E_i$.  Clearly, the allowed configurations
must satisfy $\sum_i n_i E_i \leq E$ as well as $\sum n_i = N_A$.
Assume there are a total of $V$ photon sites.

We can thus write down a formula for $\langle n_i \rangle$, the
expected number of atoms at site $i$.  There are $\binom{n_P + V -
1}{n_P}$ total possible configurations for $n_P$ photons, so the
total number of configurations of all kinds is
\begin{equation}
T = \sum_{\{n_i\}} \binom{n_P + V - 1}{n_P},
\end{equation}
where the sum is taken over atom configurations $\{n_i\}$
satisfying the constraints on total energy and total atom number.
We thus have
\begin{equation}
\langle n_j \rangle = \frac{1}{T} \sum_{\{n_i\}} n_j \binom{n_P +
V - 1}{n_P}.
\end{equation}
To proceed further, we wish to treat the photon system as a
thermal bath, so we work in the limit where $E$ and $V$ are both
very large, but $E/V$ is constant.  Then we expect most of the
energy of the system to be in the photons, so the constraint $E_A
=  \sum_i n_i E_i \leq E$ becomes unimportant, and we can neglect
terms of order $E_A/(E+V)$. Letting $x = n_P/(n_P + V - 1)$ and
$x_0 = E/(E+V)$, we have
\begin{eqnarray}
\ln \binom{n_P + V - 1}{n_P} & \approx & (n_P + V) h(x) \\
& \approx & (E+V)h(x_0) + E_A \ln x_0,
\end{eqnarray}
where we have used Stirling's formula to approximate the binomial
coefficient, and $h(x) = - x \ln x - (1-x) \ln (1-x)$. By
identifying $\beta = - \ln x_0$, we thus produce the standard
canonical ensemble for the atoms. Indeed, the procedure above
simply mimics the usual argument moving from the microcanonical
ensemble to the canonical ensemble.  Since the sum over
configurations only considers the number of atoms at each site,
the atoms behave as bosons, not as Maxwell-Boltzman particles.

We can thus immediately apply the standard results about
thermodynamics of bosons.  For instance, appropriate systems will
display Bose-Einstein condensation; we need only find an
assignment of energies to a graph which produces the correct
number of states of a given energy.  For instance, suppose we take
the positive octant of a square lattice in $3$ dimensions, and let
the energy of site $(x,y,z)$ be $x+y+z$ ($x,y,z \geq 0$).  We then
replicate the density of states for a $3$-dimensional harmonic
trap with appropriate trap frequency, and in this system, bosons
can undergo Bose-Einstein condensation~\cite{BECproof}.

The exact nature of the thermal bath is presumably not important,
and it could consist of regular quantum photons or a different
species of atoms.  While the model discussed above would be
difficult to replicate experimentally, given the variety of
possibilities, it might be possible to find some system which
could experimentally realize classical Bose-Einstein condensation.

Many further extensions are possible.  One could consider
interacting particles and could add more particle types and more
properties such as charge to try to replicate other macroscopic
quantum phenomena such as superconductivity.  One could add
velocities to make bosonic versions of lattice gas
automata~\cite{LGAreview}. However, deterministic classical boson
models face a special challenge. In a deterministic model with
distinguishable particles, two particles with the same properties
that start with the same state must retain the same state
throughout their evolution; if they were to separate, there would
be no way to determine which of the two particles would go in
which direction. We might worry that this would preclude any
interesting effects of bosonic particle statistics, but luckily
models with indistinguishable particles do not face this problem:
It is perfectly possible for two indistinguishable particles with
the same state to head off in different directions without
violating any symmetry.

One could also try to create models of classical fermions.  A
straightforward way of doing this is to simply impose a constraint
that no site can contain more than one of the particles.  This
prescription immediately reproduces the statistical mechanics of
fermions using the techniques described above for bosons, and as
before, could be produced experimentally by frequently observing
quantum fermions moving in an optical lattice. However, it is
somewhat unsatisfying from a philosophical perspective.  The
constraint causes the models to behave like fermions even if the
particles in the model are actually distinguishable.  It is
unclear if there {\em is} a true distinction between
distinguishable and indistinguishable classical fermions, but one
place to look might be in models where two or more particles can
hop at the same time (as in~\cite{cb2}). Then, for instance, if
two particles switch locations, this produces a different
configuration when the particles are distinguishable, but not if
they are indistinguishable. Unfortunately, I do not see how such
models could be realized experimentally, as a double hop in a
single time step opens the possibility for interference between
observations of the quantum systems.

In summary, I have presented some models of classical bosons,
demonstrating an error in the standard folklore that
indistinguishability is an inherently quantum property.  Indeed,
we see that indistinguishability and quantum behavior are separate
phenomena; each can exist without the other. In practice, small
objects like atoms tend to be both quantum and indistinguishable.
Larger objects have more accessible internal degrees of freedom,
so tend to lose indistinguishability. Large objects also tend to
lose quantum coherence, which is perhaps why indistinguishability
and quantum behavior have been considered in the past to be so
closely associated.

In this paper I have only examined the equilibrium behavior of the
classical models to show that they can reproduce the equilibrium
behavior of quantum indistinguishable systems.  Of course, it is
also possible to study the non-equilibrium behavior of classical
boson or classical fermion models, for instance to examine
transport properties of the models.  Indeed, it is in the realm of
dynamics that we can expect to see a difference between the
quantum and classical models, as interference can play a role in
the quantum systems but not in the classical ones.

The classical boson models offer a new perspective for
understanding macroscopic quantum phenomena.  They may even
provide an arena to make improved concrete predictions about such
phenomena: Classical systems are much easier to simulate and
analyze than quantum systems, so if classical boson (or fermion)
models can be created which replicate the major properties of
interesting systems such as high-$T_c$ superconductors, they would
provide a very useful technique for understanding those systems.
Indeed, since collective phenomena can persist for systems large
enough to experience significant decoherence, it might even be
that classical boson or fermion models are more accurate than
existing quantum models.

I would like to thank Jens Eisert, Simon Saunders, and Rafael
Sorkin for helpful conversations and Dorit Aharonov for
information on Markov processes.  This research is supported by
CIAR and NSERC of Canada.


\begin{thebibliography}{99}
\bibitem{saunders} S.~Saunders, ``On the explanation for quantum
statistics,'' Studies in the History and Philosophy of Modern
Physics, forthcoming; ``Are quantum particles objects,'' Analysis,
forthcoming.
\end{thebibliography}
\end{document}